# The nonlinear anomalous lattice elasticity associated with the high-pressure phase transition in spodumene: A high-precission static compression study

Angela Ullrich

*Institut für Geowissenschaften, Forschungsgruppe Mineralphysik*

*Universität Heidelberg, Im Neuenheimer Feld 234-236, D-69120 Heidelberg, Germany*

aullrich@min.uni-heidelberg.de

Wilfried Schranz

*Fakultät für Physik, Nichtlineare Physik, Strudlhofgasse 4, A-1090 Wien, Austria*

Ronald Miletich

*Institut für Geowissenschaften, Forschungsgruppe Mineralphysik*

# Abstract

The high-pressure behavior of the lattice elasticity of spodumene, $LiAlSi_2O_6$, was studied by static compression in a diamond-anvil cell. Investigations by means of single-crystal XRD and Raman spectroscopy within the hydrostatic limits of the pressure medium focus on the pressure ranges around ~3.2 and ~7.7 GPa, which have been reported previously to comprise two independent structural phase transitions. While our measurements confirm the well-established first-order $C2/c$-$P2_1/c$ transformation at 3.19 GPa (with 1.2% volume discontinuity and 0.81 GPa hysteresis), both unit-cell dimensions and the spectral changes observed in high-pressure Raman spectra give no evidence for structural changes related to a second phase transition. Monoclinic lattice parameters and unit-cell volumes at in total 59 different pressure points have been used to re-calculate the lattice-related properties of spontaneous strain, volume strain, and the bulk moduli as a function of pressure across the transition. A modified Landau free energy expansion in one order parameter has been developed and tested against these experimentally determined data. The Landau solution provides a much better reproduction of the observed anomalies than any equation-of-state fit to data sets truncated below and above $P_{tr}$, thus giving Landau parameters of $K_0$ = 138.3(2) GPa, $K'$ = 7.46(5), $\lambda_V$ = 33.6(2) GPa, $a$= 0.486(3), $b$ = -29.4(6) GPa and $c$ = 551(11) GPa.

# Introduction

Clinopyroxenes and their transformations have been in the focus of several experimental studies in the context of structural instabilities of various polymorphs, which are suspected to account for seismic discontinuities and anomalies of acoustic-wave propagation within the range of the Earth's upper mantle (e.g. Kung et al. 2003; 2004). Many of the seismic anomalies observed in nature have been understood as being due to compositional changes, effects of preferred orientation or changes in mineral assemblage including partial melting. Measurements of acoustic-wave velocities reveal distinct evidence for an anomalous evolution of single-crystal elastic moduli with pressure and/or temperature on approaching the critical condition boundaries for displacive transitions (e.g. Jackson et al. 2004; Kung et al., 2003; 2004; Sondergeld et al. 2006). Individual components of the tensor of elastic compliances have been found to decrease in a non-linear fashion, which gives rise for unusual softening on compression affecting not only individual lattice directions but also bulk volume compressibility. This so-called elastic softening (Carpenter and Salje 1998a; 1998b, Carpenter et al. 2000) was found to contribute over a wide pressure interval as exemplified for proper ferroelastic high-pressure transitions (Carpenter 2000; Tröster et al. 2002; Schranz et al. 2007).

High-pressure investigations in spodumene reveal distinct evidence for the occurrence of an equivalent elastic anomaly associated with the $C2/c$ to $P2_1/c$ phase transition at $P_{tr} \approx 3.2$ GPa (Arlt and Angel 2000; Sondergeld et al. 2006). Measurements of static compression by means of single-crystal XRD (Arlt and Angel 2000; Arlt et al. 2000) yield a remarkable pressure dependency of the volume-compression properties with $\partial K/\partial P = 8.9$ when fitting the volume data of the $P2_1/c$-spodumene (=high-pressure form) to a conventional Birch-Murnaghan equation of state. On the other hand, values of $\partial K/\partial P < 4$, and even negative values for individual axial compressibilities, are characteristic for low-pressure polymorphs of clinopyroxenes (Mendelson and Price 1997, Arlt and Angel 2000, Angel and Jackson 2002, Gatta 2003, Webb and Jackson 1993, Angel and Hugh-Jones 1994). Therefore, it was not surprising that Sondergeld et al. (2006) could verify the expected discontinuous evolution of single-crystal elastic moduli at the pressure-induced transition at ~3.2 GPa. Direct evidence for elastic softening is given from ultrasonic interferometry measurements as obtained from in-situ investigations in a multi-anvil press, where compressional wave velocities in single-crystal samples indicate discontinuities at the transition pressure. A conversion of measured $v_P$ to elastic constants, which correspond to the $a$, $b$ and $c^*$ axial directions, clearly reveals an abrupt discontinuity marking reproducible the onset of the transition. In particular for the $c^*$ direction this reveals a softening of the axial elastic constant from ~280 GPa below $P_{tr}$ to ~150 GPa at the critical transition pressure. A Landau free energy expansion was developed within the scope of this study in order to describe the observed elastic anomaly and to approach the deviations of lattice-related quantities from smooth evolution of elastic stiffness of the lattice. All the latter consideration by Sondergeld et al. (2006) related to lattice properties, such as lattice parameters, unit-cell volumes, spontaneous strain, but also densities used for converting the acoustic-wave velocities to elastic constants, have been based on the compression data given by Arlt and Angel (2000). Undoubtedly, these results obtained from high-pressure single-crystal diffraction measurements are of excellent quality, but the restricted number of data points in direct proximity to the critical transition pressure does not provide the sufficient resolution to reproduce the anomaly properly. In addition the components of spontaneous strain accompanying the structural phase transition reveal unusual patterns not scaling with $Q^2$ ($Q$ = order parameter). Hence either contributions from higher-order strain/order parameter couplings or a pressure dependency of the coupling coefficients has to be considered. All findings suggest to sample more data points of lattice parameters in small pressure increments close to $P_{tr}$. Moreover, Pommier et al. (2003) claim a second phase transition to occur at about 7.7 GPa, as argued from distinct spectral changes in high-pressure Raman spectra. Within the scope of this study we re-investigated the critical pressure ranges at 3.2 and 7.7 GPa by means of both single-crystal X-ray diffraction and Raman investigations. One focus was to get further insights in the possible existence of the second transition, but our major focus was the precise determination of the lattice dimensions in very fine pressure steps across the established first-order transition at 3.2 GPa. Based on in total 59 data points, lattice-related quantities such as spontaneous strain, volume strain and bulk moduli were determined and a modified approach of Landau theory was applied to the new experimental data.

# Experimental and analytical procedures

## Sample material and high-pressure conditions

All sample crystals used in this study were fragments from a several cm big single crystal specimen of kunzite (Springer collection, of the Institut für Geowissenschaften, University of Heidelberg, inventory number S462, unknown locality, most probably Afghanistan). Optically clear cleavage fragments of about 100 x 80 x 50 µm in size were chosen for the high-pressure experiments in diamond anvil cells (DACs). Pressures up to 9.2 GPa have been generated by opposed-anvil ETH-type DACs (Miletich et al. 2000), a re-development from BGI-type four-pin DAC (Allan et al. 1996). DACs were equipped with X-ray transparent beryllium backing plates with a 12° optical port suitable for spectroscopy applications. Type Ia-diamond anvils of standard geometry (height: 1.4 mm, table-face diameter: 2.8 mm) with culets of 600 µm have been used for all measurements. Sample chambers were provided by stainless steel gaskets preindented to about 90 µm with bore holes of 200-250 µm. A 4:1 methanol-ethanol mixture was used as pressure transmitting medium providing hydrostatic pressure conditions at the samples. In addition to the sample single crystal a quartz single crystal was added serving as pressure calibrant for the X-ray measurements and rubies for pressure determination during Raman spectroscopy.

## Raman scattering

Raman spectra have been collected within the hydrostatic regime of the pressure medium at 14 pressures between 1 and 9.0 GPa. A LabRam HR800 UV spectrometer equipped with an OLYMPUS BXFM-ILHS optical microscope, automated x-y stage, a grating with 1800 grooves per millimeter, a Peltier-cooled CCD detector and a 100x objective (numerical aperture 0.90) was used for data collection. Excitation source was the 632.8 nm line of a He-Ne laser and the spectra were recorded in backscattered geometry. Lateral resolution was estimated to be smaller than 1.5 μm, wavenumber accuracy was 0.5 $cm^{-1}$ and the spectral resolution was 1 $cm^{-1}$. Peak positions have been obtained by fitting Lorentzian polynomials to the measured bands. Pressure was monitored using the He-Ne laser of the Raman spectrometer for excitation of the characteristic ruby fluorescence (Mao et al. 1978) and the error for deriving the pressure is estimated to be about 0.05 GPa in maximum.

### Single crystal X-ray diffraction

Lattice parameters have been measured between $10^{-4}$ and 9.3 GPa. Profiles of Bragg intensities were recorded on a Huber 5042 four-circle diffractometer with a point detector, horizontal and vertical diffracted-beam slits, and using unmonochromatized Mo-radiation from a conventional sealed-tube X-ray source operated at 50kV and 32mA. Constrained lattice parameters of spodumene and the quartz crystal have been refined at each pressure step from the positions of 20 Bragg reflections. Corrections were carried out by applying the method of eight-position diffracted-beam centering (King and Finger 1979), which has been used to eliminate errors in peak positions due to sample displacement. Peak scanning, centering and lattice vector least-squares refinements from the corrected positional angles were conducted using the SINGLE04 software (Ralph and Finger 1982, Angel et al. 2000a). Pressures at each pressure step were determined from the refined unit-cell volumes of quartz using the equation of state parameters given by Angel et al. (1997).

## Results and Discussion

### Raman spectroscopy

Selected characteristic Raman spectra, one of the low-P spodumene at 3.41, GPa, one above $P_c$ at 3.58 GPa, and one in the range of the suspected second transition at 7.07 GPa are given in Figure 1. The corresponding band positions fitted from all 14 spectra are listed in Table 1, their evolution with pressure is shown in Figure 2.

The observed band positions are consistent with those reported by Pommier et al. (2003) and have been labeled following their notation. For the $C2/c$ phase we observe 18 bands, which is more compared to the number of observed bands in the study of Pommier et al. (2003). According to factor group analysis, there exist theoretically up to 30 Raman-active modes (14 $A_g$ and 16 $B_g$. All additional peaks reported here have been labeled $n_1$, $n_2$, $n_3$ and $n_4$. (see Figure 1). From theory there are more modes which could not be detected probably due to peak degeneracy, low intensity and background contributions.

Between 3.4 and 3.6 GPa an increase from 18 to 26 bands clearly marks the onset of the phase transition from $C2/c$ to $P2_1/c$ space group symmetry. According to Pommier et al. (2003) the individual vibrational modes could be assigned to Si-O stretching vibrations of the non-bridging Si-O bonds (between 800 and 1200$cm^{-1}$), the modes between 650 and 800$cm^{-1}$ to stretching vibrations of the bridging Si-O vibrations, and the modes between 425 and 650$cm^{-1}$ to Si-O bending vibrations. The increase in the number of observed modes at higher wavenumbers has been attributed to the appearance of two distinct silicate chains in the $P2_1/c$ structure. As noticed by Pommier (2003) the singlet between 650 and 800cm$^{-1}$ remains a singlet although from studies of other pyroxenes (Ross et al. 1999) a doublet would have been expected in the $P2_1/c$ structure. Changes of the modes in the low wavenumber region (50 and 425cm$^{-1}$, bending and stretching of the M1-O and M2-O bonds) can be ascribed to the breaking of the $C_2$ point group symmetry at the M1 and M2 site.

It has been noticed by Pommier (2003) that some modes ($\upsilon_7$, $\upsilon_8$, $\upsilon_{11}$ and $\upsilon_{19}$) disappear and one peak ($\upsilon_{12}$) appears at about 7 GPa. On basis of this observation the onset of a second phase transition has been suspected. In our study these peaks have been observed even up to pressures higher than 7.7 GPa. Only marginal changes in peak intensities have been observed for $\upsilon_3$, $\upsilon_4$, $\upsilon_7$ and $\upsilon_{10}$ at 7.1 GPa which in our opinion does not justify to assume a structural change. The evolution of both band position and intensity in the spectra do not give strong indications for a second transition or, a least, does not account for structural changes of the same magnitude as those which are observable at 3.2 GPa.

## Lattice compression

The results of the single-crystal diffraction measurements of lattice parameters (*a*, *b*, *c* and $\beta$) and unit-cell volumes at 59 different pressures is summarized in Table 2. The variations of the unit-cell volume and the individual lattice parameters with pressure are plotted in Figure 3. Apart from only very small differences in unit-cell dimensions, most likely due to the use of a different sample, the results are consistent with the data previously reported by Arlt and Angel (2000). The first-order phase transition from space group $C2/c$ to $P2_1/c$ is accompanied by a significant volume reduction of 1.2 % and has been found to be reversible with an apparent but small hysteresis. During compression the transition takes place between 3.21 and 3.23 GPa, the equivalent volume discontinuity was observed between 3.19 and 3.18 GPa on decompression. With the careful measurement in very fine pressure steps around $P_{tr}$ the outlying data point reported by Arlt and Angel (2000) at 3.19 GPa turned out to be not reproducible (see Figure 3). However, the previously observed indications for the occurrence of softening-related anomalies could be reconfirmed. Equivalent fits of the truncated *P*-*V*, *P*-*a*, *P*-*b* and *P*-*c* data sets to a Birch-Murnagahn equation-of-state (Table 3 Angel et al 2000b) confirm the findings of Arlt and Angel (2000) and reveal higher compressibilities for the high-pressure phase than for the low-pressure phase. In addition, the fits resemble significantly larger $\partial K/\partial P$ dependencies for all compressional moduli of the high-pressure lattice.

## Spontaneous strain and Landau theory

Spontaneous strain, which is indicative for lattice distortion across related to the symmetry changes, was calculated according to Carpenter et al. (1998a) Figure 4 shows the evolution of the individual components of the spontaneous strain tensor varying with pressure. In consistency with the relatively small discontinuity observed in the *b*-axis direction, the discontinuity in $e_{22}$ at $P_{tr}$ is relatively small compared to the strains $e_{11}$, $e_{33}$ and $e_{13}$ arising from the discontinuous changes in the *a*-axis, the *c*-axis and the $\beta$-angle. Since the transition involves no change in point group symmetry, all strain components are non-symmetry breaking, hence are expected to evolve according to $e_{ij} \propto Q^2 \propto (P_{tr}-P)^{0.5}$. Whereas $e_{33}$ and the shear strain $e_{13}$ evolve according to the expected relation the strain components $e_{11}$ and $e_{22}$ reveal almost no change above $P_{tr}$.

In order to describe the deviation from normal compression behavior associated with the phase transition, a Landau free energy expansion in the driving order parameter $Q$ has been determined using the ISOTROPY software (Stokes et al. 2007). The active representation of the $C2/c$-$P2_1/c$ transition is $Y_2^-$. The form of the full Landau expansion to sixth order in $Q$, including coupling terms between order parameter and strain components $e_i$, was recently worked out by Sondergeld et al. (2006). In the following we intend to describe the anomalous pressure dependence of the bulk modulus in the vicinity of the phase transition. For this purpose one can ignore the complexities introduced by considering the full form of the order parameter-strain couplings of the type $\lambda_i Q^2 e_i$ as well as the full elastic energy $1/2 C^o_{ij} e_i e_j$. Following Sondergeld et al. (2006) we use a simplified Landau expansion including the bulk modulus $K^0$ of the $C2/c$ phase at $P = 0$ and the volume strain $e_V$ (Sondergeld et al. 2006):

$$F(Q,e_V) = \frac{1}{2}AQ^2 + \frac{1}{4}BQ^4 + \frac{1}{6}CQ^6 + \lambda_V e_V Q^2 + \frac{1}{2}K^0 e_V^2 \quad (1)$$

In Sondergeld et al. (2006) the second-order term in the Landau expansion is assumed to be linearly dependent on the hydrostatic pressure, i.e. $a(P-P_c)$. This "mixing" of the conjugated variables stress = $-P$ and strain = $e_V$ causes some problems in the further treatment of the problem. We therefore proceed in a different way, by assuming our second order term in (1) to be constant or temperature dependent, i.e. $a(T-T_c)$, but independent on pressure. Minimizing Eq.(1) with respect to the volume strain

$$\frac{\partial F(Q,e_V)}{\partial e_V} = \sigma = -P = \lambda_V Q^2 + K^0 e_V \quad (2)$$

one obtains

$$e_V = -\frac{P}{K^0} - \frac{\lambda_V}{K^0} Q^2 \quad (3)$$

Performing the Legendre Transformation F($e$)→G(P)

$$G(P) = F(e) + VPe \quad (4)$$

one obtains with Eq´s.(3) and (1) the following form for the Gibbs free energy

$$G(Q,P) = -\frac{1}{2}a(P-P_c)Q^2 + \frac{1}{4}b^*Q^4 + \frac{1}{6}cQ^6 - \frac{P^2}{2K^0} \quad (5)$$

with the following relations between the Landau coefficients

$$a = \frac{2\lambda_V}{K^0} > 0, \quad b^* = B - \frac{2\lambda_V^2}{K^0}, \quad c = C \quad \text{and} \quad P_c = \frac{AK^0}{2\lambda_V} = \frac{A}{a} \quad (6)$$

Note that Eq.(5) has the same form as was used by Sondergeld et al. (2006) for the pure order parameter part which is called $G(L)$ in their Eq.(5). But due to our different procedure, we can here relate the coefficient $a$ to the parameters $\lambda_V$ and $K^0$, whereas in Sondergeld, et al. (2006) $a$ is completely free. Using the fit parameters $\lambda_V$ = 34.4 GPa and $K^0$ = 144.2 GPa one obtains $a$ = 0.48 in very good agreement with the value (0.5) found by Sondergeld et al. (2006) (see Table 3 of their work).

Minimizing Eq.(5) with respect to $Q$ yields for the equilibrium value of the order parameter

$$Q^2 = \frac{2}{3}(Q_{tr})^2\left\{1 + \left[1 - \frac{3}{4}\frac{(P-P_c)}{(P_{tr}-P_c)}\right]^{\frac{1}{2}}\right\} \tag{7}$$

Since the phase transition is first order in character, one obtains three characteristic pressures, which we denote here as $P_1$, $P_{tr}$ and $P_c$. $P_{tr}$ can be identified with the measured transition pressure, i.e. $P_{tr}$ = 3.19 GPa. The other two values determine the stability limits of the high- and low symmetry phases, where the following well known relations hold:

$$P_{tr} = P_c - \frac{3b^{*2}}{16ac} \quad \text{and} \quad P_1 = P_c - \frac{4b^{*2}}{16ac} \tag{8}$$

Using the fit parameters of Table 3 of Sondergeld, et al. (2006), i.e. $b^*$ = -41 GPa, $a$ = 0.5 and $c$ = 769 GPa one obtains for the region of coexistence for both phases $P_c$-$P_1$ = 1.09 GPa. This defines the maximum pressure range of the observable hysteresis. Due to the first-order nature of the phase transition the order parameter displays a volume discontinuity at $P_{tr}$, given by the relation

$$Q^2(P_{tr}) = -\frac{3b^*}{4c} \tag{9}$$

Again, in conformity with Sondergeld et al. (2006), $Q^2(P_{tr})$ has been chosen to be 0.04.

The pressure-dependent evolution of the bulk modulus can then be calculated using the relation

$$K = -V_0\frac{\partial P}{\partial V} = -\left(\frac{\partial \varepsilon_V}{\partial P}\right)^{-1} \tag{10}$$

Using Eq´s.(3) and (10) one obtains

$$K(P) = \frac{K^0}{1 + 2\lambda_V Q(P)\frac{\partial Q(P)}{\partial P}} \tag{11}$$

With Eq.(7) the pressure dependence of the bulk modulus can be written as

$$K(P) = \frac{K^0}{1 - \frac{2\lambda_V^2}{K^0 b^*}\left[1 - \frac{3}{4}\frac{(P-P_c)}{(P_{tr}-P_c)}\right]^{\frac{1}{2}}} \tag{12}$$

with $K^0 = K_0 + K'P$ and $K_0$ denoting the zero-pressure bulk modulus and $K' = \partial K/\partial P$.

Figures 5a and b show the pressure dependencies of the total volume strain and the bulk modulus calculated from the lattice parameter data. The detailed data collection in the proximity of the phase transition provides the exact form of the curvature in the pressure dependent evolution of the bulk modulus. Landau parameters have been obtained from fits according to Eq’s (3) and (12) to the variations of volume strain and bulk modulus simultaneously (Table 5). The Landau parameters determined by Sondergeld et al. (2006) and the EOS parameters of the $C2/c$ phase have been used as starting parameters. Based on equations (6), (8) and (9) we related the parameters $a$, $b^*$ and $P_c$ to the parameters $\lambda_V$, $K_0$, $c$ and $P_{tr}$. With the critical transition pressure $P_{tr}$ fixed at 3.19 GPa, only three fit parameters are completely free, which minimizes the uncertainties determined for the parameters. Compared to Sondergeld et al. (2006) a slightly smaller hysteresis of 0.81 GPa has been found. The Landau solution is in good agreement with the experimental data, reproducing the discontinuous evolution of the volume strain in the range of the phase transition. However deviation from experimental data has been found towards higher pressures. The general form of the pressure-dependent evolution of the bulk modulus is reproduced successfully revealing compressional changes in a nonlinear fashion between 3.2 and about 6 GPa. But the extend of the anomaly close to $P_{tr}$ deviates from the predictions of Landau theory; immediately above $P_{tr}$ the experimentally determined bulk moduli are much lower than Landau theory predicts. Whereas the observed bulk moduli tend to approach zero at $P_{tr}$, the lowest values obtained for the bulk modulus following the Landau solution gives approximately 83 GPa. Although Landau theory provides a good reproduction of the general form of the elastic anomalies associated with the phase transition, it does not account for the large volume discontinuity associated with the strongly first-order phase transitions. In addition the inconsistency between experimental data and Landau solution, which has been observed towards high pressures in the volume strain (Figure 5a), is also visible in the evolution of the bulk modulus (Figure 5b). These discrepancies can be ascribed to the treatment of strains as infinitesimal. It is difficult to include pressure effects beyond infinitesimal approximation in the 'traditional' Landau theory. However, at very high pressures the assumption of 'linear elasticity' is not sufficient and one has to deal with finite strains. But considering finite strains and introducing such nonlinear assumptions makes the formulation of Landau theory far more complicated which this is beyond the scope of this study.

# Conclusion

This work presents a detailed analysis of the elastic anomalies associated with the first-order $C2/c$-$P2_1/c$ phase transition taking place in spodumene at $P_{tr}$ = 3.19 GPa. Lattice parameters determined in very fine pressure increments provide the evolution of the components of spontaneous strain, the volume strain and the bulk modulus in the range of the phase transition, yielding the exact form of the observed softening by approaching $P_{tr}$ from high pressures. Unit cell data and Raman spectroscopic data reveal continuous compression above $P_{tr}$ and the existence of a second phase transition at 7.7 GPa appears to be very unlikely. A Landau free energy expansion has been derived in the driving order parameter and in comparison with a truncated equation of state fit the Landau solution provides much better agreement with the general form of observed anomalies. Contrary to Sondergeld et al. (2006) who assumed the second-order term in the Landau expansion to be linearly dependent on the hydrostatic pressure, i.e. "mixing" the conjugated variable stress and strain, we assumed our second-order term to be constant or temperature dependent, i.e. $a(T$-$T_c)$, but independent on pressure. In addition by minimizing the number of fit parameters and by fitting the evolution of the volume strain and the bulk modulus simultaneously, the uncertainties afflicting the fit parameters have been minimized. Compared to the study of Sondergeld et al. (2006), revealing an drop in $K_0$ to about 120 GPa at $P_{tr}$, our modified Landau expansion accounts better for the volume discontinuity associated with the strongly first-order character of the phase transition. However the expansion still does not fully account for the extend of the discontinuity. In addition deviations between experimental and calculated data have been observed towards high pressures, which have been ascribed to the treatment of spontaneous strain as infinitesimal.

## Acknowledgements

This research was supported within the scope of the project grant MI 605/2-2 of the Deutsche Forschungsgemeinschaft (DFG). Special thanks are also due to the GSI, Gesellschaft für Schwerionenforschung mbH, Darmstadt (Materials Research Group) for providing the possibility to use the Raman spectrometer for this study. Support by the Austrian FWF (P19284-N20) is gratefully acknowledged.

## Figure Captions

Figure 1: Variation of Raman Spectra with pressure showing distinct evidence for the onset of the $C2/c$-$P2_1/c$ phase transition between 3.4 and 3.6 GPa (peaks have been labelled according to Pommier et al. (2003).

Figure 2: Plot of the wavenumber peak positions changing as a function of pressure.

Figure 3: Evolution of the unit cell parameters of spodumene with pressure showing a first order phase transition from $C2/c$ to $P2_1/c$. (a) Variation of the unit-cell volume and (b)-(d) variations of the individual lattice parameters (data collected in this study is shown by closed symbols, data collected by Arlt et al. (2000) by open symbols, solid lines represent Murnaghan fits using the parameters listed in Table 3).

Figure 4: Variation of the individual components of the spontaneous strain arising due to the $C2/c$-$P2_1/c$ phase transition.

Figure 5: Pressure dependent changes of the volume strain (a) and the bulk modulus (b) in spodumene. Values for the bulk modulus have been calculated from the change of the unit-cell volume with pressure according to $-(1/2)(V_1+V_2)(P_1-P_2)/V_1-V_2)$; the solid lines symbolizes the Landau solutions using the parameters given in Table 4, the dashed line the background strain and the dotted line shows the Murnaghan equation-of-state determined for the $P2_1/c$ phase.

| No. | 1.76 | 3.20 | 3.41 | 3.58 | 3.80 | 3.97 | 4.83 | 7.07 | 8.19 | 8.97 |
|---|---|---|---|---|---|---|---|---|---|---|
| | | | | | | | | 117 | 112 | 114 |
| | 130 | 129 | 129 | 138 | 138 | | 139 | 140 | 141 | 142 |
| | 189 | 190 | 190 | 185 | 185 | | 186 | 189 | 191 | 191 |
| $\upsilon_1$ | | | | 212 | 213 | 214 | 216 | 220 | 223 | 224 |
| $\upsilon_2$ | 231 | 232 | 232 | 230 | 230 | 231 | 234 | 240 | 243 | 244 |
| $\upsilon_3$ | | | | 249 | 249 | 250 | 253 | 259 | 264 | 266 |
| $\upsilon_4$ | 253 | 255 | 255 | 262 | 262 | 263 | 265 | 271 | 274 | 275 |
| $\upsilon_5$ | | | | 287 | 288 | 289 | 291 | 298 | 302 | 304 |
| $\upsilon_6$ | 302 | 306 | 306 | | | | | | | |
| $\upsilon_7$ | | | | 319 | 320 | 320 | 323 | 329 | 331 | 331 |
| $\upsilon_8$ | 334 | 339 | | 338 | 339 | 342 | 347 | 359 | 363 | 365 |
| $\upsilon_9$ | 358 | 361 | 362 | | | | | 374 | 379 | 384 |
| $\upsilon_{10}$ | | | | 373 | 374 | 375 | 378 | 387 | 391 | 393 |
| $\upsilon_{11}$ | | 386 | 388 | 389 | 390 | 392 | 397 | | | |
| $\upsilon_{12}$ | | | | | | | | | 423 | 422 |
| $\upsilon_{13}$ | 397 | 400 | 400 | | | | | | | |
| $\upsilon_{14}$ | | | | 411 | 412 | 414 | 417 | 428 | 436 | 441 |
| $\upsilon_{15}$ | 423 | 427 | 427 | | | | | | | |
| $\upsilon_{16}$ | | | | 442 | 443 | 444 | 446 | 452 | 456 | 458 |
| $\upsilon_{17}$ | 450 | 457 | 458 | 455 | 456 | 458 | 461 | 469 | 475 | 477 |
| $\upsilon_{18}$ | | | | 496 | 497 | 498 | 501 | 509 | 514 | 516 |
| $\upsilon_{19}$ | 527 | 530 | 531 | 528 | 529 | 529 | 531 | 535 | 538 | 539 |
| $\upsilon_{20}$ | | | | 541 | 542 | 542 | 544 | 550 | 552 | |
| | 550 | 553 | 551 | | | | | | | |
| | | | | 574 | 574 | 573 | 577 | 583 | 588 | 590 |
| $\upsilon_{21}$ | 590 | 593 | 593 | 597 | 597 | 598 | 599 | 605 | 608 | 610 |
| $\upsilon_{22}$ | 713 | 718 | 719 | 721 | 721 | 722 | 725 | 731 | 735 | 737 |
| $\upsilon_{23}$ | 890 | 897 | 897 | 897 | 899 | 900 | 904 | 904 | 908 | 914 |
| $\upsilon_{24}$ | 986 | 993 | 994 | 989 | 990 | 992 | 995 | 1005 | 1009 | 1011 |
| $\upsilon_{25}$ | | | | 1022 | 1022 | 1023 | 1025 | 1031 | 1035 | 1036 |
| $\upsilon_{26}$ | 1027 | 1033 | 1034 | | | | | | | |
| $\upsilon_{27}$ | | | | 1040 | 1039 | 1039 | 1042 | 1052 | 1052 | 1057 |
| $\upsilon_{28}$ | 1037 | 1047 | 1051 | 1057 | 1056 | 1056 | 1060 | 1070 | 1073 | 1078 |
| $\upsilon_{29}$ | 1083 | 1090 | 1091 | 1091 | 1092 | 1093 | 1097 | 1107 | 1114 | 1118 |

**Table 1** Raman shifts obtained from fits of Lorentzian polynomials to the measured data.

| $V_{Qtz}$ (Å$^3$) | $P$ | $a$ (Å) | $b$ (Å) | $c$ (Å) | $\beta$ | $V$ (Å$^3$) |
|---|---|---|---|---|---|---|
| 112.994(10) | 0.000(3) | 9.4649(6) | 8.3934(6) | 5.2190(8) | 110.146(8) | 389.25(7) |
| 112.994(10) | 0.000(3) | 9.4664(4) | 8.3920(4) | 5.2195(6) | 110.155(5) | 389.20(5) |
| 110.071(8) | 1.052(3) | 9.4402(5) | 8.3701(7) | 5.2085(3) | 110.113(4) | 386.45(4) |
| 110.064(7) | 1.055(3) | 9.4403(3) | 8.3705(6) | 5.2086(2) | 110.114(3) | 386.48(3) |
| 109.947(11) | 1.101(4) | 9.4389(4) | 8.3702(7) | 5.2083(3) | 110.117(4) | 386.38(4) |
| 109.081(7) | 1.454(3) | 9.4297(3) | 8.3622(5) | 5.2044(2) | 110.093(3) | 385.40(3) |
| 108.779(8) | 1.581(4) | 9.4269(4) | 8.3605(4) | 5.2033(3) | 110.076(3) | 385.17(3) |
| 108.242(10) | 1.813(5) | 9.4209(4) | 8.3560(4) | 5.2013(3) | 110.066(3) | 384.60(3) |
| 107.479(6) | 2.157(3) | 9.4135(4) | 8.3466(7) | 5.1970(2) | 110.064(3) | 383.55(3) |
| 106.802(6) | 2.475(3) | 9.4051(3) | 8.3405(5) | 5.1938(2) | 110.044(2) | 382.74(3) |
| 106.754(6) | 2.498(3) | 9.4054(3) | 8.3395(6) | 5.1939(2) | 110.046(3) | 382.71(3) |
| 106.648(6) | 2.550(3) | 9.4032(4) | 8.3400(7) | 5.1934(3) | 110.042(3) | 382.62(3) |
| 106.589(5) | 2.578(2) | 9.4023(3) | 8.3375(5) | 5.1928(2) | 110.039(3) | 382.43(3) |
| 106.476(6) | 2.633(3) | 9.4024(4) | 8.3371(8) | 5.1929(3) | 110.041(3) | 382.42(4) |
| 106.425(7) | 2.659(3) | 9.4018(3) | 8.3361(6) | 5.1925(2) | 110.040(3) | 382.32(3) |
| 106.094(8) | 2.823(4) | 9.3972(3) | 8.3338(5) | 5.1905(2) | 110.028(3) | 381.91(3) |
| 106.063(8) | 2.838(4) | 9.3977(4) | 8.3339(5) | 5.1905(2) | 110.028(3) | 381.93(3) |
| 105.935(8) | 2.903(4) | 9.3957(3) | 8.3336(5) | 5.1897(2) | 110.025(3) | 381.79(3) |
| 105.866(8) | 2.938(4) | 9.3946(3) | 8.3306(6) | 5.1891(2) | 110.019(3) | 381.58(3) |
| 105.646(8) | 3.050(4) | 9.3939(9) | 8.3293(15) | 5.1889(6) | 110.002(8) | 381.51(8) |
| 105.604(8) | 3.072(4) | 9.3917(4) | 8.3289(6) | 5.1879(2) | 110.019(3) | 381.29(3) |
| 105.563(8) | 3.093(4) | 9.3921(4) | 8.3291(8) | 5.1879(5) | 110.018(4) | 381.31(5) |
| 105.496(8) | 3.128(4) | 9.3907(4) | 8.3277(6) | 5.1880(3) | 110.019(4) | 381.20(4) |
| 105.450(8) | 3.152(4) | 9.3906(4) | 8.3265(6) | 5.1877(3) | 110.014(3) | 381.13(3) |
| 105.419(5) | 3.168(3) | 9.3887(5) | 8.3285(8) | 5.1872(4) | 110.011(4) | 381.12(4) |
| 105.414(10) | 3.171(5) | 9.3888(7) | 8.3279(12) | 5.1859(6) | 110.008(7) | 381.01(7) |
| 105.405(7) | 3.175(4) | 9.3907(4) | 8.3276(6) | 5.1880(3) | 110.019(4) | 381.20(4) |
| 105.405(7) | 3.175(4) | 9.3894(3) | 8.3289(6) | 5.1873(3) | 110.011(3) | 381.17(3) |
| 105.386(6) | 3.186(3) | 9.3898(3) | 8.3269(6) | 5.1871(2) | 110.012(3) | 381.08(3) |
| 105.370(5) | 3.194(3) | 9.3890(3) | 8.3258(6) | 5.1869(2) | 110.015(3) | 380.98(3) |
| 105.369(5) | 3.194(3) | 9.3889(3) | 8.3253(7) | 5.1872(3) | 110.014(4) | 380.98(4) |
| 105.343(7) | 3.208(3) | 9.3886(3) | 8.3260(6) | 5.1868(2) | 110.012(3) | 380.97(3) |
| 105.374(5) | 3.192(3) | 9.3139(8) | 8.3668(8) | 5.1150(9) | 109.282(11) | 376.24(13) |
| 105.360(8) | 3.199(4) | 9.3157(3) | 8.3663(6) | 5.1152(3) | 109.286(3) | 376.30(3) |
| 105.353(7) | 3.203(3) | 9.3149(3) | 8.3677(7) | 5.1146(3) | 109.281(3) | 376.29(4) |
| 105.333(6) | 3.213(3) | 9.3142(4) | 8.3671(7) | 5.1142(3) | 109.279(4) | 376.21(4) |
| 105.306(7) | 3.227(4) | 9.3146(3) | 8.3664(5) | 5.1143(3) | 109.280(3) | 376.20(3) |
| 105.300(6) | 3.231(3) | 9.3142(3) | 8.3668(6) | 5.1141(2) | 109.276(3) | 376.20(3) |
| 105.224(6) | 3.271(3) | 9.3134(3) | 8.3669(6) | 5.1136(3) | 109.278(3) | 376.13(3) |
| 105.215(7) | 3.275(3) | 9.3129(3) | 8.3642(6) | 5.1136(3) | 109.275(4) | 375.99(4) |
| 105.191(7) | 3.288(4) | 9.3131(3) | 8.3651(5) | 5.1132(3) | 109.274(3) | 376.01(3) |
| 105.144(7) | 3.313(3) | 9.3130(4) | 8.3650(8) | 5.1120(4) | 109.267(4) | 375.94(5) |
| 105.110(6) | 3.331(3) | 9.3124(3) | 8.3651(5) | 5.1122(2) | 109.268(3) | 375.93(3) |
| 105.100(6) | 3.336(3) | 9.3118(3) | 8.3638(7) | 5.1117(3) | 109.270(3) | 375.80(4) |
| 105.093(7) | 3.340(4) | 9.3105(3) | 8.3640(6) | 5.1115(3) | 109.266(3) | 375.75(3) |
| 105.016(9) | 3.381(5) | 9.3096(3) | 8.3627(6) | 5.1107(3) | 109.263(3) | 375.60(4) |
| 104.718(14) | 3.542(9) | 9.3062(9) | 8.3610(10) | 5.1059(7) | 109.253(8) | 375.07(7) |
| 104.35(2) | 3.745(11) | 9.3007(7) | 8.3542(9) | 5.1022(4) | 109.204(6) | 374.38(5) |
| 103.909(16) | 3.994(9) | 9.2966(9) | 8.3483(10) | 5.0974(7) | 109.209(9) | 373.59(8) |
| 102.855(8) | 4.618(5) | 9.2800(7) | 8.3395(9) | 5.0859(5) | 109.136(6) | 371.86(5) |
| 102.465(14) | 4.859(9) | 9.2757(7) | 8.3324(9) | 5.0822(6) | 109.141(8) | 371.08(7) |
| 101.753(13) | 5.314(8) | 9.2671(4) | 8.3267(7) | 5.0759(4) | 109.092(4) | 370.11(4) |
| 101.396(10) | 5.549(7) | 9.2612(8) | 8.3219(9) | 5.0702(5) | 109.065(7) | 369.33(5) |
| 100.949(15) | 5.852(10) | 9.2545(6) | 8.3174(11) | 5.0659(6) | 109.069(7) | 368.55(7) |
| 99.876(7) | 6.613(5) | 9.2407(5) | 8.3042(5) | 5.0552(3) | 109.005(4) | 366.77(4) |
| 98.400(9) | 7.743(7) | 9.2201(4) | 8.2855(7) | 5.0392(4) | 108.953(4) | 364.09(4) |
| 97.462(9) | 8.517(8) | 9.2066(3) | 8.2735(6) | 5.0297(3) | 108.921(3) | 362.42(3) |
| 96.959(9) | 8.950(7) | 9.2000(4) | 8.2662(7) | 5.0240(4) | 108.905(4) | 361.46(4) |
| 96.637(9) | 9.235(8) | 9.1982(4) | 8.2598(8) | 5.0207(5) | 108.897(5) | 360.89(5) |

**Table 2** Unit-cell parameters of spodumene varying with pressure

| | spodumene (HT) $C2/c$ | spodumene $P2_1/c$ |
|---|---|---|
| $a_0$ (Å) | 9.4661(2) | 9.419(2) |
| $K_0$ (GPa) | 124.0(4) | 76(3) |
| $K'$ | 4(fixed) | 13.5(5) |
| $b_0$ (Å) | 8.3931(2) | 8.439(2) |
| $K_0$ (GPa) | 127.3(5) | 113(5) |
| $K'$ | 4(fixed) | 7.6(9) |
| $c_0$ (Å) | 5.2196(17) | 5.209(3) |
| $K_0$ (GPa) | 163.2(1.1) | 42(2) |
| $K'$ | 4(fixed) | 11.3(4) |
| $V_0$ (Å$^3$) | 389.29(2) | 389.1(2) |
| $K_0$ (GPa) | 142.5(5) | 79.2(2) |
| $K'$ | 4(fixed) | 11.0(4) |

**Table 3** Equation-of-state parameters of spodumene

| | |
|---|---|
| $K_0$ | 138.3(2) GPa |
| $K'$ | 7.46(5) |
| $\lambda_V$ | 33.6(2) GPa |
| $P_{tr}$ | 3.19(fix) GPa |
| $P_c$ | 3.795(13) GPa |
| $a$ | 0.486(3) |
| $b^*$ | -29.4(6) GPa |
| $c$ | 551(11) GPa |

**Table 4** Values of Landau parameters

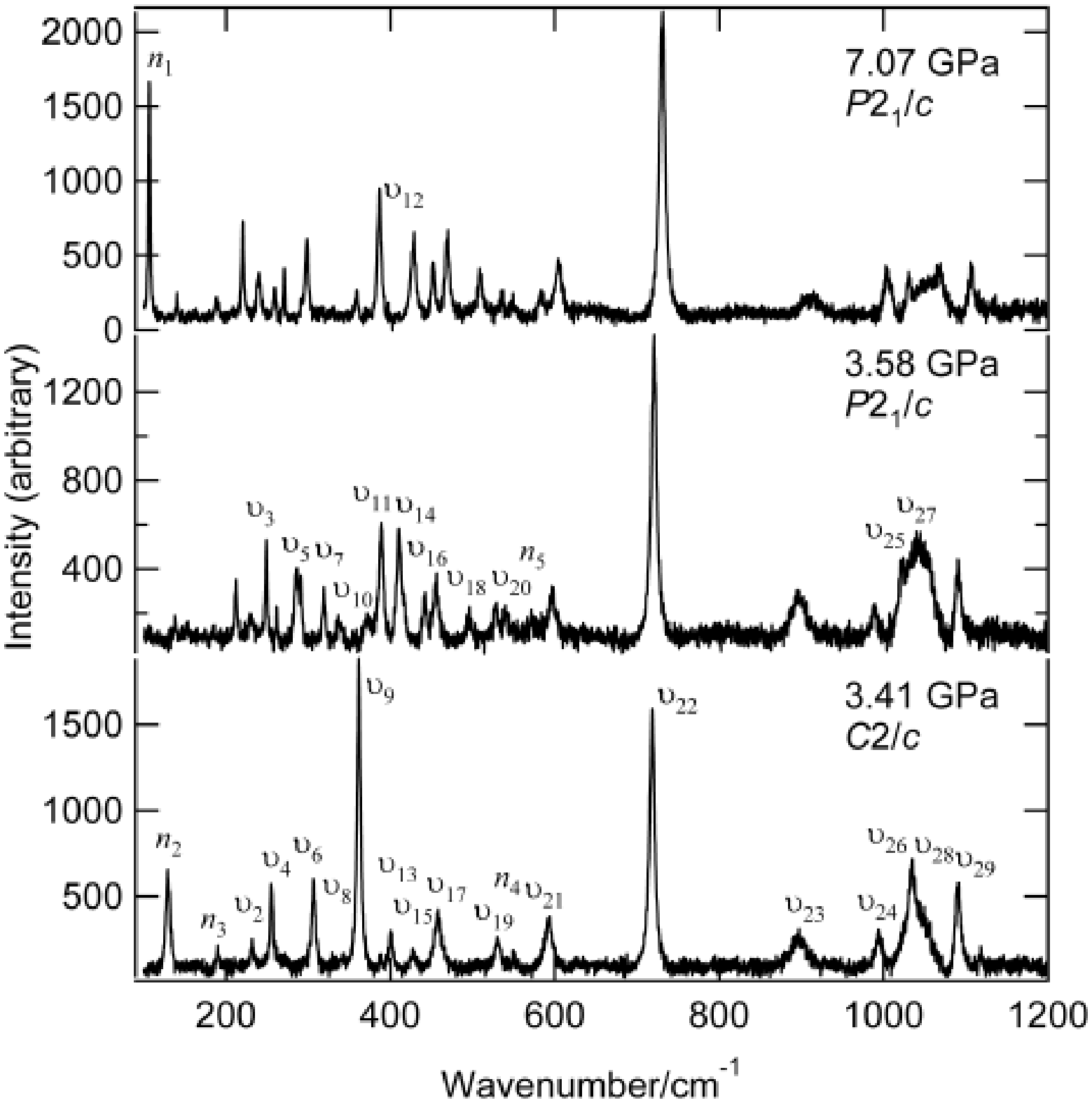

7.07 GPa
P2₁/c
3.58 GPa
P2₁/c
3.41 GPa
C2/c
Intensity (arbitrary)
Wavenumber/cm-1

**Figure 1**

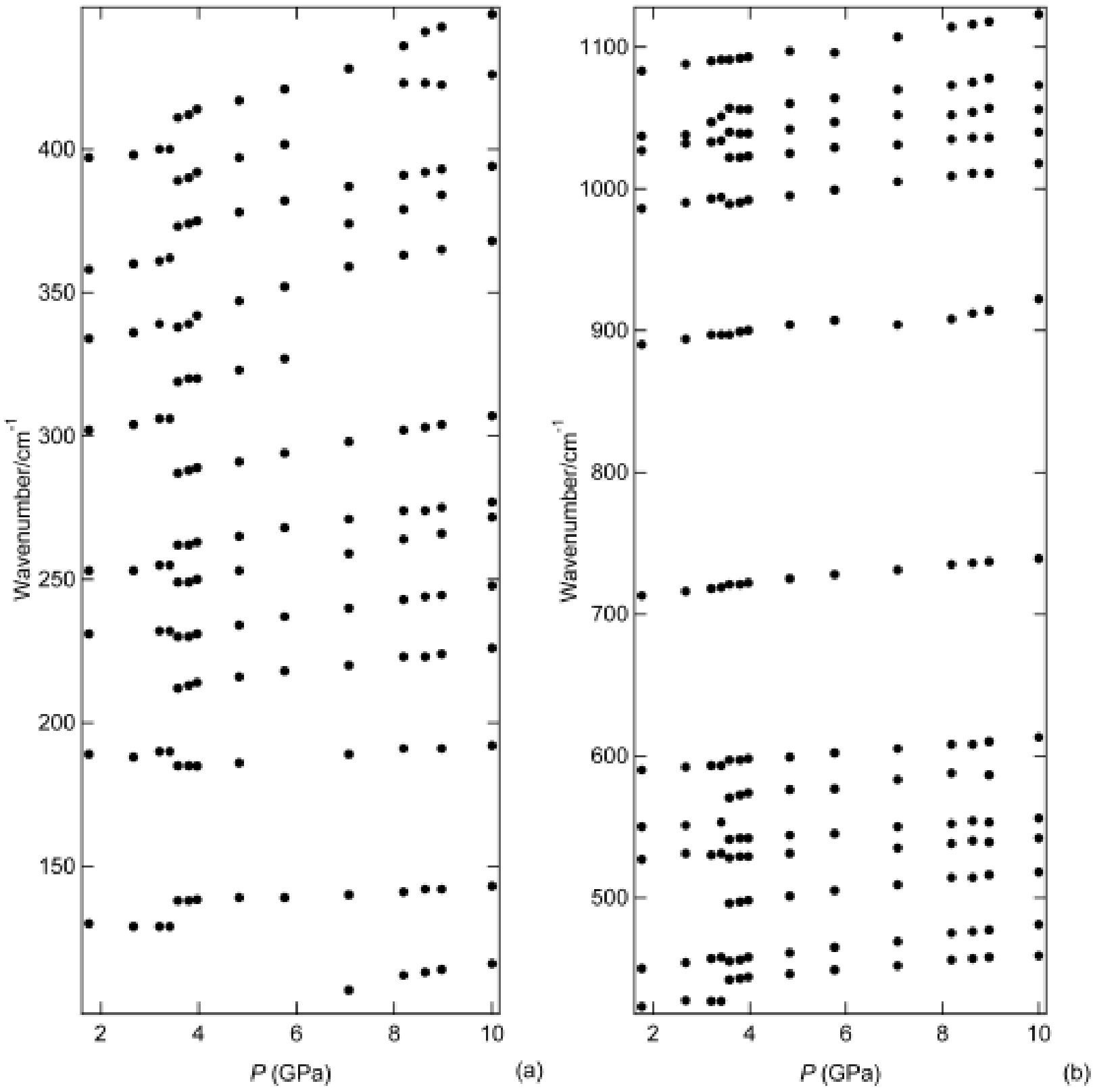

**Figure 2**

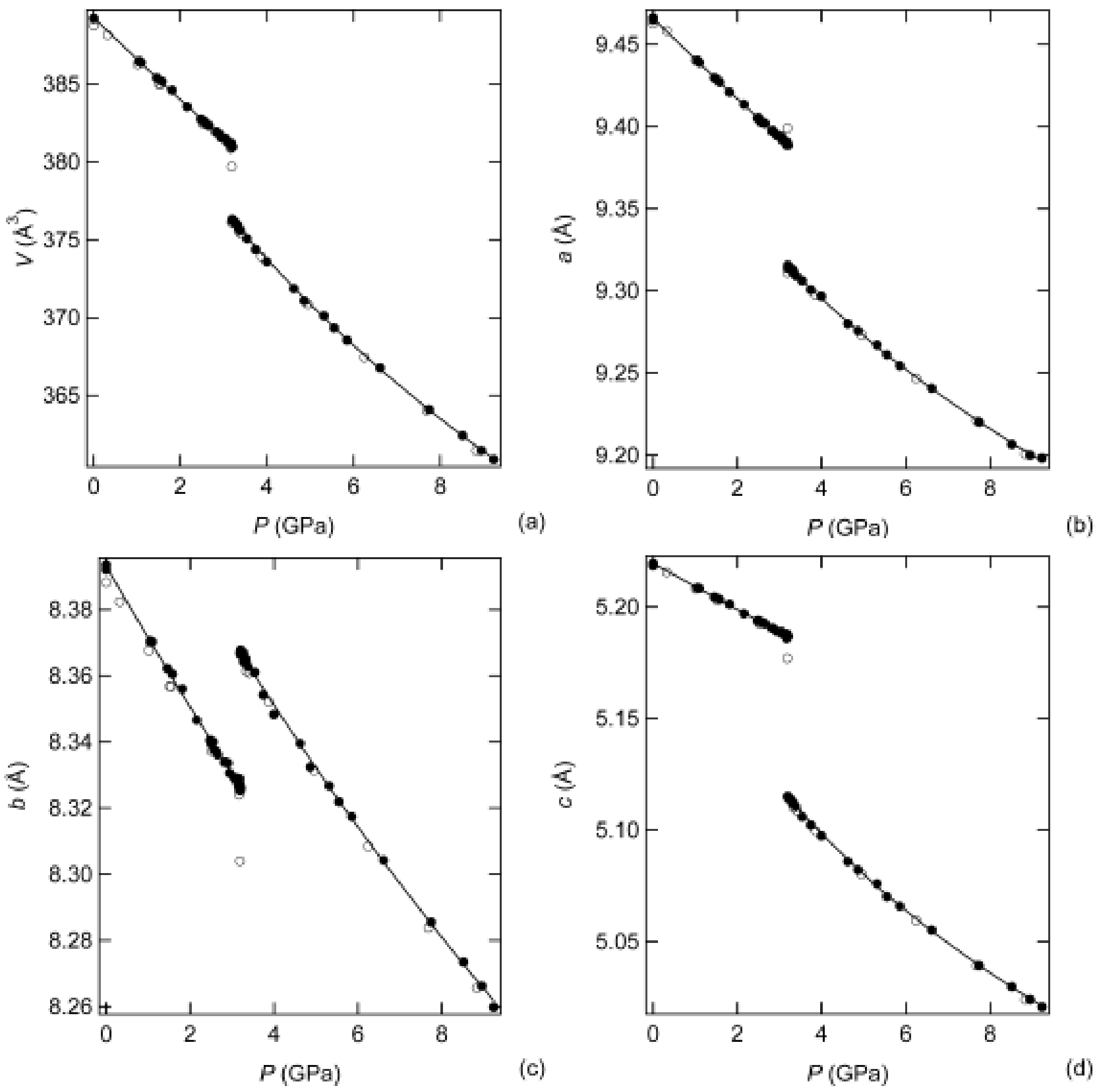

**Figure 3**

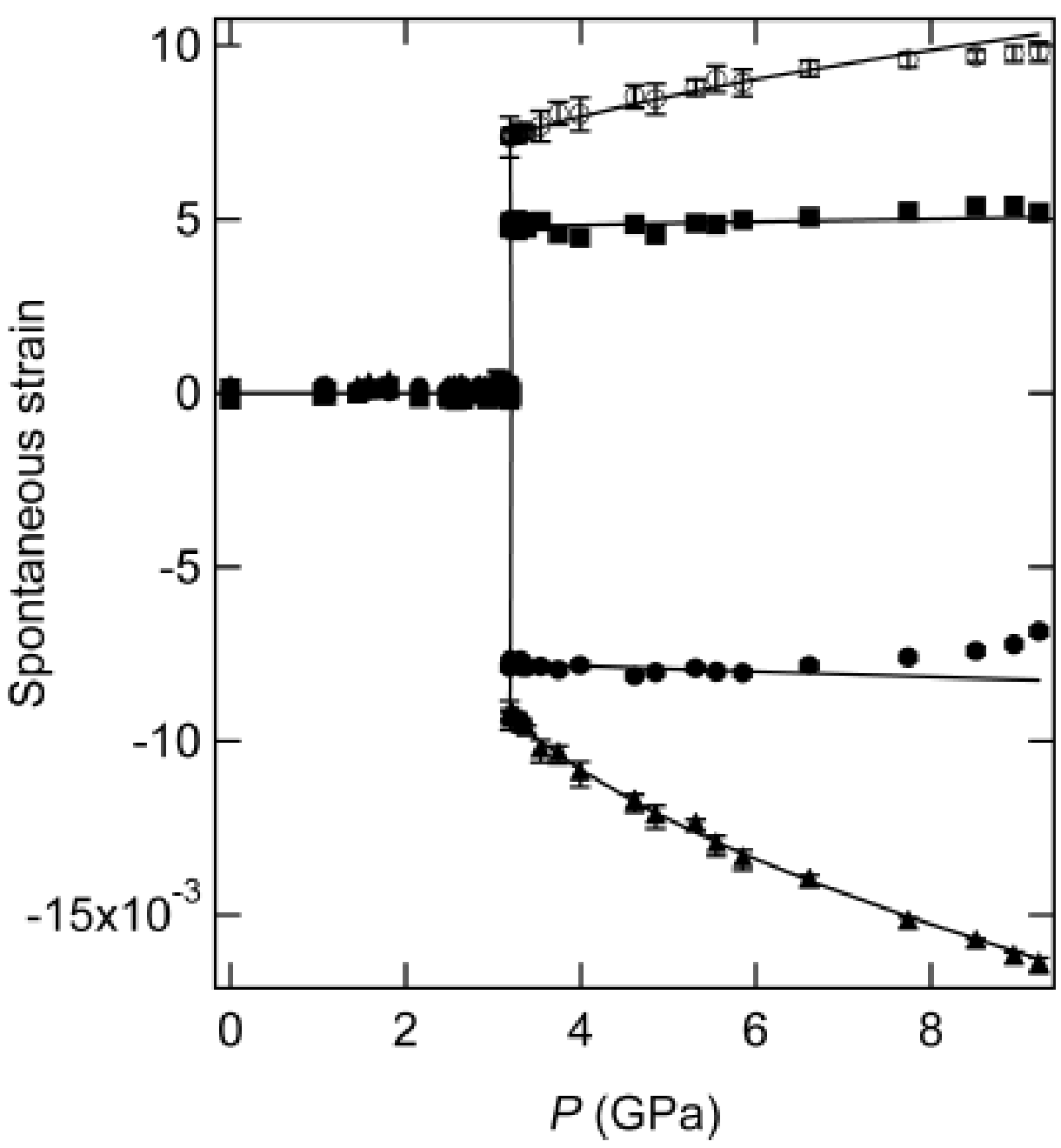

**Figure 4**

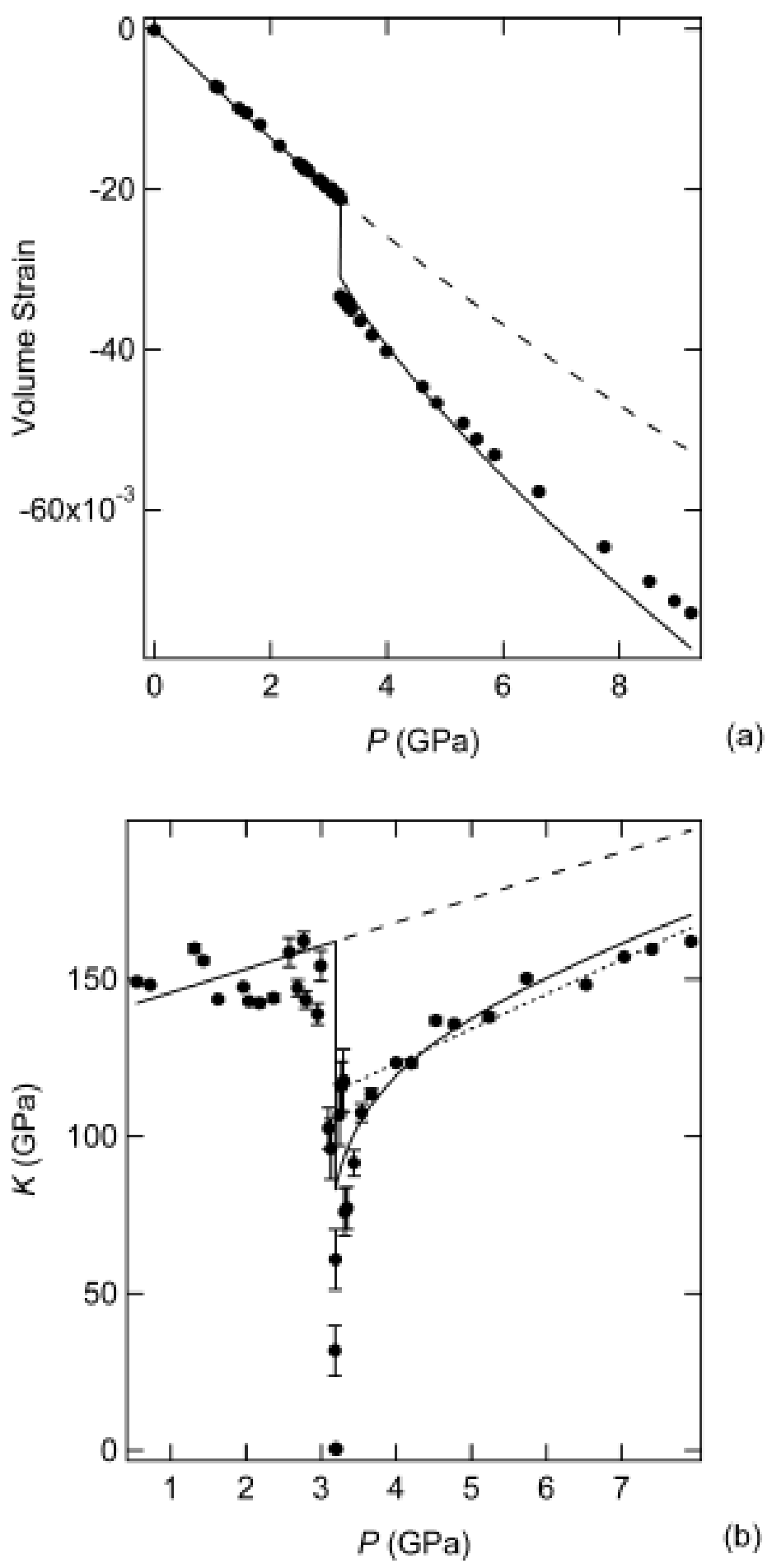

**Figure 5**